\documentclass[twocolumn,final]{IEEEtran}
 \usepackage{url,tabularx,verbatim,paralist}
\usepackage{graphicx}			% Include figure files
\usepackage{dcolumn}			% Align table columns on decimal point
\usepackage{bm}					% bold math
\usepackage{float}			% Include float files         https://www.ctan.org/pkg/float?lang=en
\usepackage{amsmath,amsfonts}	% popular packages from the American Mathematical Society
\usepackage{url}					% https://www.ctan.org/pkg/url
\usepackage{setspace}		% Sets spacing    https://www.ctan.org/pkg/setspace
\usepackage[autostyle]{csquotes}
\usepackage{paralist}
\usepackage{subcaption}
\usepackage{amsthm}
\theoremstyle{plain}
\newtheorem *{thm*}{Sklar's Theorem}
\newtheorem *{def*}{Definition}
\newtheorem *{defvine*}{Definition}
\newtheorem *{defcopula*}{Definition}

\begin{document}

\title{Dependence Structure Analysis Of Meta-level Metrics in YouTube Videos: A Vine Copula Approach}      % don't need 2 lines, example to show linebreak \\

\author{Vikram Krishnamurthy,  {\em Fellow, IEEE}  \and \hspace{1cm} Yan Duan
\thanks{Vikram Krishnamurthy (corresponding author) is
 with the  School of Electrical and Computer
Engineering, and Cornell Tech,  Cornell University, New York 
(email:  vikramk@cornell.edu). This research was partially supported by an Army Research office grant and the Schmidt Sciences.
Yan Duan is with the  University of British Columbia,  Canada.  (email: yduan1@ece.ubc.ca).
}}

\maketitle

\begin{abstract} 
This paper uses vine copula to analyze the multivariate statistical dependence in a massive YouTube dataset consisting of 6 million videos over 25 thousand channels. Specifically we study the statistical dependency of 7 YouTube meta-level metrics: view count, number of likes, number of comments, length of video title, number of subscribers, click rates, and average percentage watching. Dependency parameters such as the Kendall's tau and tail dependence coefficients are computed to evaluate the pair-wise dependence of these meta-level metrics. The vine copula model yields several interesting dependency structures. We show that view count and number of likes' are in the central position of the dependence structure. Conditioned on these two metrics, the other five meta-level metrics are virtually independent of each other. Also, Sports, Gaming, Fashion, Comedy videos have similar dependence structure to each other, while the News category exhibits a strong tail dependence. We also study Granger causality effects and upload dynamics and their impact on view count. Our findings provide a useful understanding of user engagement in YouTube.
\end{abstract}

%  End of title page -------------------------------------------------------------------------------------------------------------------------------- %

% \linenumbers				% start linenumbers here

\section{\label{sec:introduction} Introduction}

Since its establishment in 2005, YouTube has over 1 billion registered users, who spend millions of hours and generate billions of views everyday; more than 300 hours of video content are uploaded every minute~\cite{stephens2015big}. 
While YouTube is  a social media site, is is also a social networking site.  Classical online social networks (OSNs) are dominated
by   user-user interactions. However YouTube is unique in that the interaction between users  is more sophisticated since it includes video content--that is, the interaction follows: users $\rightarrow$  content $\rightarrow$  other users. 

The interaction between users in the YouTube social network is incentivized using the posted videos. 
%In this way it is not merely the interest or homophily between users that promote user-user interaction, but also the content of the videos that governs the social interactions.
These interactions  include:
\begin{compactenum}
\item Commenting on users' videos and  commenting on other users comments which is very similar to users interaction on blog posting sites such as Twitter.
\item Subscribing to YouTube channels provides a method of forming relationships between users. 
\item Users can directly comment on a YouTube channel. % without the need to only interact when a video is posted. 
\item Users can also interact by embedding videos from another users channel directly into their own channel to promote exposure or form communities of users. 
\end{compactenum}

In this paper we view YouTube as a big-data time series. We statistically analyze a massive YouTube dataset of more than 6 million videos over 25 thousand channels to understand how various YouTube meta-level metrics (such as view count, subscriber count,  number of likes, etc) affect each other.
YouTube is a useful source of data: with efficient web crawlers, meta-level metrics data such as \enquote{number of views} can be collected for millions of videos. Careful  analysis of YouTube data can reflect users' preference and lead to useful outcomes in optimal caching~\cite{hoiles2015adaptive}, recommendation systems~\cite{zhou2010impact,davidson2010YouTube} and targeted advertising~\cite{zhang2011improving}. %Many papers dealing with statistical analysis of YouTube data focus on studying correlations between user preferences and video characteristics. 
For example \cite{hoiles2015adaptive} studies how popularity prediction of YouTube can be used to design efficient caching algorithms in 5G telecommunication systems.

\subsection{Context. Copulas}
With the rapid growth of YouTube data sets, there is strong motivation for adequately understanding and modeling the dependencies  present
in the resulting multivariate random processes. 
The modeling of a multivariate distribution function can be decomposed into  considering the marginal distributions and then determining the underlying dependence structure, the so-called copula. One of the most popular copula classes, especially for high- dimensional data, are vine copulas.

In this paper we focus on the correlations amongst YouTube video metrics and their dependence relationships, which are determined using vine copula models. 
 Vine copula models \cite{durante2010copula}   have several attractive features: since they do not impose constraints on marginal distributions, they can capture unusual correlations such as extreme co-movements~\cite{trivedi2007copula}. Furthermore, with vine copula, one can separate the task of estimating marginal distributions from that of estimating dependence between random variables. A short review of bivariate copulas and vine copula model is provided in the Appendix.

\subsection{\label{sec:main_results} Main Results}

%\subsection{\label{sec:main_results} Main Results}
 This paper uses vine copula to determine the multivariate dependency structures in  a YouTube dataset consisting of more than 6 million videos over 25 thousand channels. The dataset consists of the daily samples of meta-level video metrics from April, 2007 to May, 2015.   
Specifically we consider 7 YouTube meta-level metrics: view count, number of likes, number of comments,  length of video title, number of subscribers, click rates, and average
percentage watching.
%Based on the vine copula analysis of the YouTube dataset, we unravel several interesting dependence structures: 
%In particular, we specify the dependence structure  a graphical vine copula structure, and pair-wise measures of dependence are computed based on the vine copula model. 

The vine copula model is useful for visualizing the interdependencies of the YouTube metrics across different channels.  Also, the Kendall's Tau yields the relative importance of these metrics.	
Based on the vine copula analysis of the YouTube dataset, we conclude several important facts regarding dependence structures (Sec.\ref{sec:results} describes these dependencies in more detail) of YouTube metrics.
\begin{compactenum}
\item We found that 
 \enquote{number of views}\footnote{The  view count is a key metric of the measure of popularity or ``user engagement" of a video and the metric by which YouTube pays the content providers} and \enquote{number of likes} are in the central position of the dependence structure. Conditioned on these two
 metrics, the other five YouTube metrics are virtually independent of each other
(see Figure \ref{fig:five}). 
So for example, conditioned on the number of likes, the number of comments (negative or positive) is statistically independent of the number
of subscribers (which at first sight is somewhat counterintuitive).
Surprisingly, this dependency structure is true for all categories of videos that we considered, namely,
gaming, sports, fashion  and comedy.

\item Regarding the various categories of videos;  \enquote{News} videos have stronger statistical inter-dependence when their values are large (i.e. positive extreme co-movements), 
 For \enquote{Gaming} videos, \enquote{annotation clicks rate} is less dependent on number of views compared to other categories,
Users watching \enquote{Comedy} videos are more likely to give comments.
For a video with certain popularity, users are more likely to finish a video in \enquote{Fashion} category than in other categories.

\item Finally, given the dependency structure from the vine copula, we dig further into the dynamics of YouTube.
We use Granger causality to determine the causal relation between viewcounts and subscribers for channels in YouTube.
We also study the upload dynamics of YouTube and find he interesting property that for popular gaming YouTube channels with a dominant upload schedule, deviating from the schedule increases the views and the comment counts of the channel.

\end{compactenum}

The vine copula models in this paper 
 are of interest in interactive advertising \cite{PC15,KAB16}: based on the meta-level metrics, which channels to advertise on?
Also, knowledge of the dependency structure allows YouTube content providers to maximize  the number of views. The interaction between users in the YouTube is incentivized using the posted videos. In addition to the social incentives, YouTube also gives monetary incentives to promote users increasing their popularity. As more users view and interact with a users video or channel, YouTube will pay the user proportional to the advertisement exposure on the users channel. Therefore, users not only maximize exposure to increase their social popularity, but also for monetary gain. 
Finally,  in 5G wireless networks \cite{hoiles2015adaptive},  adaptive caching of content is crucial: the vine copula gives a tractable model to determine what videos to cache.

\subsection{\label{sec:literature_review} Literature Review}
Copulas are now used in wide range of areas including finance, econometrics, biology and medicine \cite{Joe}. 
Research on YouTube video data analysis started as early as 2007. Pioneering works include: meta-data statistics/social network analysis~\cite{cha2007tube,cheng2008statistics}, and proposed methods to improve quality of service (QoS)~\cite{cheng2009accelerating,hossfeld2013internet}. For the dependence analysis on video meta-level metrics, earlier works found that YouTube videos have clear small-world characteristics and thus strong correlations with each other, and such knowledge can lead to more efficient video caching  and redistribution~\cite{cheng2008statistics,krishnappa2015cache}. However, to our best knowledge, no statistical analysis has been conducted on the dependence among the key YouTube video metrics.

This paper is organized as follows. In Sec.\ref{sec:analysis}, we discuss the main results of this paper: summary of the YouTube data statistics, details for the four steps of dependence analysis using vine copula, estimated parameters, and interpretations of the results. In Sec.\ref{sec:simulation}, we validate the vine copula model using the White goodness-of-fit test. Finally, Sec.\ref{sec:conclusions} offers the concluding remarks. In the Appendix, we review the Sklar's theorem and bivariate Archimedean copulas, which lead to the construction of the vine copula model.

\section{A Short Outline of Vine Copulas} \label{sec:shortreview}

Since its introduction by Sklar in 1959~\cite{nelsen2013introduction}, copulas are  a useful multivariate model in various fields \cite{bouye2000copulas,subramanian2011fusion,zhang2006bivariate}.

\begin{defcopula*}
A (Borel-measurable) function $C:[0,1]^n \rightarrow [0,1]$ is a copula, if the following properties hold:
\begin{compactitem}
\item for each $i \in \{1,2,...,n \}$, $C(\textbf{u}) = u_i$ when all components of $\textbf{u}$ are equal to $1$, except the $i^{th}$ item that is equal to $u_i \in [0,1]$.  

\item for all $\textbf{u},\textbf{v} \in [0,1]^n$ such that $\forall i \in \{1,2,...,n\}$, $u_i \leq v_i$, $C(\textbf{u}) \leq C(\textbf{v})$

\item $C$ is $n$-decreasing.
\end{compactitem}
\end{defcopula*}

A fundamental  result involving copula is Sklar's Theorem.

\label{sec:sklar}
\begin{thm*}
Let $F \in \mathcal{F}(F_1,...,F_n)$ be an $n$-dimensional cumulative distribution function with continuous marginal distribution functions $F_1,...F_n$. Then there exists a unique copula $C$ with uniform marginals: $C \in \mathcal{F}(\mathcal{U},...,\mathcal{U})$, such that for all $x_1,. . . , x_n$ in $(-\infty,\infty)$
\begin{equation}
\label{eqn: hey1}
%\label{eqn:copula}
F(x_1,...,x_n)=C(F_1(x_1),...,F_n(x_n)).
\end{equation}
\end{thm*}
 Note that copula function $C$ is a $[0,1]^n \rightarrow [0,1]$ mapping. Therefore, in copula applications, the raw data is required to be transformed into uniformly distributed data, or the so-called \enquote{copula data} by using probability integral transformation (PIT).
Sklar's theorem states that as long as one can estimate marginal distributions for metrics of interests, it is possible to find a function (or \enquote{copula}) to properly model their multivariate distribution. %However, the theorem doesn't specify what kind of function is the best candidate for a certain scenario we want to model. The approach that most copula applications adopt is to start from existing \enquote{templates}: indeed there are already various parametric copulas available for different scenarios. 

 Bedford and Cooke~\cite{bedford2001probability} proposed the so-called \enquote{Regular Vine} (or simply \enquote{R-Vine}) graphical structure,\footnote{A  vine is a graphical tool for labeling constraints in high-dimensional probability distributions.} hence the name \enquote{vine-copula}. \enquote{Vine} refers to a nested set of tree structures, where the edges of the $i^{th}$ tree are the nodes of the ${i+1}^{th}$ tree. 
 In copula applications, the trees are used to represent the dependence structure of multiple variables: each edge represents the bivariate copula connecting one pair of marginals; see appendix for an illustrative example.

Formally a regular vine structure is defined as follows (see Appendix for background):
\begin{defvine*}
A graphical structure $V$ is a regular vine of $m$ elements if:
\begin{enumerate}
\item $V = (T_1,...,T_{m-1})$ and all trees are connected
\item The first tree $T_1$ has node set $N_1 = {1,...,m}$ and edge set $E_1$; then for the next trees $T_i$, $i \in {2,...m-1}$, $T_i$ has the node set $N_i = E_{i-1}$
\item Proximity condition: If two nodes are connected by an edge in the ${i+1}^{th}$ tree, then the two edges in $i^{th}$ tree corresponding to these two nodes share a node.
\end{enumerate}

\end{defvine*}

The following steps (see appendix for details)  construct a regular vine copula (R-vine) specification to a dataset:
\begin{enumerate}
\item Construct the R-Vine structure by choosing an appropriate unconditional and conditional pair of metrics to use for vine copula model.

\item Choose a bivariate copula family for each pair selected in step (1).

\item Estimate the corresponding parameters for each bivariate copula.
\end{enumerate}

%%%%%%%%%%%%%%%%%%
\section{\label{sec:analysis} Dependence Structure of YouTube using Vine Copula}

In this section, we use the vine copula model to unravel the dependence structure of  a massive YouTube data set.

\subsection{\label{sec:data} YouTube Dataset}
%This paper uses the dataset provided by BBTV. 
%This section provides a brief introduction to the dataset and summarizes the statistics of the dataset. 
The dataset contains daily samples of metadata of YouTube videos  from April, 2007 to May, 2015, and has a size of around $200$ gigabytes. 
%The dataset contains around $6$ million videos spread over $25$ thousand channels. 
Table~\ref{tab:dataset:summary}  summarizes the dataset. 
%The dataset contains channels which are popular and contain large number of videos. 
\begin{table}[h!]
	\centering
	\caption{Dataset summary}
	\begin{tabular}{c|c}
		\hline
		Videos &  $6$ million\\
		Channels &  $26$ thousand\\
		Average number of videos (per channel)& 250 \\
		Average age of videos & 275 days\\
		Average number of views  (per video) & $10$ thousand \\
		\hline
	\end{tabular}
	\label{tab:dataset:summary}
\end{table}

The data-set comprises
seven meta-level metrics and  five categories of YouTube videos.
The 7 meta-level metrics of YouTube:
 \begin{compactitem} 
 \item Number of views
 \item Number of likes
 \item Number of comments
 \item Video title length
 \item Number of subscribers
 \item Annotation click rates
 \item Average percentage of watching
\end{compactitem}
The YouTube videos we consider belong to five categories:
 \begin{compactitem} 
 \item Sports
 \item Fashion
 \item Gaming
 \item Comedy
 \item News
\end{compactitem}

 %In Sec.\ref{sec:marginal}, the marginal distribution for each metric is empirically estimated, Kolmogorov-Smirnov test is applied to confirm that raw data of seven YouTube video metrics is transformed into uniformly distributed copula data. In Sec.\ref{sec:dependence}, we introduce the \enquote{Kendall's tau}, which is used by the vine copula model to measure the pair-wise dependence of YouTube video metrics. In Sec.\ref{sec:results}, the main results regarding the dependence structure of YouTube video metrics are discussed in detail.

 For each category, 6000 samples including data of seven metrics are chosen from the YouTube dataset, which has different number of videos for different categories. For the reader's convenience, the abbreviations of long metrics names are also provided. All of above information is summarized in Table \ref{tab:categories}.

\begin{table*} \centering
%\begin{ruledtabular}
\begin{tabular}{cccccccc}
Number of Videos& Categories & Metrics & Abbreviation\\
\hline
85657 &News & Number of Views & Views\\
\hline
90586 &Fashion & Number of Likes & Likes\\
\hline
60438&Sports & Number of Comments & Comm\\
\hline
431803&Gaming & Video Title Length & TitLen\\
\hline
58820&Comedy & Number of Subscribers & Subs\\
\hline
Null & Null & Annotation Clicks & Anno\\
\end{tabular}
%\end{ruledtabular}
\caption{YouTube Video Data Categories, Metrics and Abbreviations} \label{tab:categories}
\end{table*}

%%%%%%%%%%%%

\begin{table*}[p]

  \caption{Summary of YouTube data}
\label{tab:summary_of_Youtube data}

 \begin{subtable}[h]{1\textwidth}
  \caption*{(A): News}
 
  \centering
% \begin{ruledtabular}
\begin{tabular}{cccccccc}
Stats & Views  & Likes & Comm & TitLen & Subs & Anno & AvgPer\\
Min & 1 & 0 &0 &4 &0 &0 &0\\ % inserting body of the table
\hline
Max & 5775400 & 12111 & 9035 & 120 & 9793 & 35.78 & 99.03\\
\hline
Median & 715 & 2 & 1 & 50 & 0 & 0 & 56.40\\
\hline
Mean & 25540 & 124.50 & 71.43 & 50.66 & 26.62 & 0.27 & 50.77\\
\hline
Kurtosis & 9.59 & 11.44 & 7.83 & 0.23 & 9.81 & 10.41 & 0.74\\
\hline
Skewness & 1.61 & 1.04 & 1.27 & 0.21 & 2.15 & 1.25 & -0.43\\
\end{tabular}
%\end{ruledtabular}
\end{subtable}

\begin{subtable}[h]{1\textwidth}
  \centering
  \caption*{(B): Comedy}

%   \begin{ruledtabular}
\begin{tabular}{cccccccc}
Stats & Views  & Likes & Comm & TitLen & Subs & Anno & AvgPer\\
Min & 1 & 0 &0 &4 &0 &0 &0\\ % inserting body of the table
Min & 1 & 0 &0 &1 &0 &0 &0\\ % inserting body of the table
\hline
Max & 4355984 & 134103 & 22932 & 120 & 23338 & 408.764 & 99.04\\
\hline
Median & 1385 & 19 & 9 & 36 & 1.00 & 0.0 & 40.17\\
\hline
Mean & 50264 & 487.7 & 95.64 & 38.69 & 54.36 & 2.096 & 40.43\\
\hline
Kurtosis & 14.32 & 13.46 & 11.82 & 0.041 & 13.30 & 5.82 & 0.042\\
\hline
Skewness & 5.05 & 3.18 & 2.91 & 0.096 & 3.68 & 2.00 & 0.57\\
\end{tabular}
%\end{ruledtabular}
\end{subtable}

\begin{subtable}[h]{1\textwidth}
  \centering
  \caption*{(C): Sports}

%     \begin{ruledtabular}
\begin{tabular}{cccccccc}
Stats & Views  & Likes & Comm & TitLen & Subs & Anno & AvgPer\\
Min & 0 & 0 &0 &4 &0 &0 &0\\ % inserting body of the table
\hline
Max & 5516714 & 12361 & 5963 & 120 & 1589 & 109.1267 & 92.07\\
\hline
Median & 1772 & 11 & 6 & 54 & 1.0 & 0.0 & 39.71\\
\hline
Mean & 17298 & 110.3 & 42.29 & 55.25 & 13.6 & 1.198 & 40.34\\
\hline
Kurtosis & 12.15 & 12.30 & 6.14 & 0.21 & 2.30 & 1.26 & 0.05\\
\hline
Skewness & 3.10 & 1.52 & 2.14 & 0.10 & 1.30 & 0.90 & 0.066\\
\end{tabular}
%\end{ruledtabular}
\end{subtable}

\begin{subtable}[h]{1\textwidth}
  \centering
  \caption*{(D): Fashion}
  
%\begin{ruledtabular}
\begin{tabular}{cccccccc}
Stats & Views  & Likes & Comm & TitLen & Subs & Anno & AvgPer\\
\hline % inserts single horizontal line
Min & 1 & 0 &0 &3 &0 &0 &0\\ % inserting body of the table
\hline
Max & 6900910 & 22846 & 5094 & 120 & 7553 & 176.8490 & 98.53\\
\hline
Median & 1287 & 14 & 8 & 45 & 2.00 & 0.0 & 28.73\\
\hline
Mean & 30666 & 128 & 36.18 & 45.05 & 36.76 & 1.0688 & 29.70\\
\hline
Kurtosis & 4.44 & 5.63 & 4.85 & 0.037 & 4.70 & 3.65 & 0.03\\
\hline
Skewness & 1.83 & 1.87 & 1.79 & 0.062 & 1.30 & 1.62 & 0.049\\
\end{tabular}
%\end{ruledtabular}
\end{subtable}

%\end{table}

%\begin{table}[h!]

\begin{subtable}[h]{1\textwidth}
  \centering
  \caption*{(E): Gaming}

%  \begin{ruledtabular}
\begin{tabular}{cccccccc}
Stats & Views  & Likes & Comm & TitLen & Subs & Anno & AvgPer\\
\hline % inserts single horizontal line
Min & 0 & 0 &0 &2 &0 &0 &0\\ % inserting body of the table
\hline
Max & 3233762 & 28314 & 9731 & 120 & 7405 & 91.4822 & 98.36\\
\hline
Median & 341 & 6 & 4 & 54 & 0.00 & 0.0 & 26.72\\
\hline
Mean & 7380 & 121 & 39.22 & 54.37 & 12.02 & 0.6450 & 30.21\\
\hline
Kurtosis & 12.16 & 4.94 & 5.99 & 0.027 & 13.17 & 2.52 & 0.063\\
\hline
Skewness & 3.21 & 1.96 & 2.29 & 0.021 & 3.82 & 1.40 & 0.11\\
\end{tabular}
\end{subtable}
%\end{ruledtabular}

\end{table*}

 Table \ref{tab:summary_of_Youtube data} conducts  a preliminary statistical analysis on the dataset in terms of mean, median, kurtosis and skewness.
It can be concluded that apart from \enquote{title length} and \enquote{average percentage of watching}, the other five metrics of YouTube videos have very high kurtosis and skewness, which imply heavy tails and high asymmetry of those distributions. Therefore, these five metrics of YouTube videos are not normally distributed. It is this property that motivates the use of vine copula:  the vine copula does not require the marginals to be normally distributed, it serves as a useful multivariate model for the dependence analysis of YouTube dataset.

\subsection{\label{sec:marginal} Modelling Marginal Distributions}
   
The marginal empirical  distribution of each YouTube metric can be estimated from the data:
\begin{equation}
F^{N}(x)=\frac{1}{N+1}\sum^{N}_{i=1}I(X_i \leq x).
\end{equation}
where $N$ is the number of data samples, $X_i$ is the $i^{th}$ data point, while $I(\cdot)$ is the indicator function.
The probability integral transform (PIT)  is then applied to the estimated marginal distribution to obtain uniformly distributed samples in [0,1], which is referred as \enquote{copula data} and required by the vine copula model. In order to test whether the copula data is uniformly distributed in $[0,1]$, the Kolmogorov-Smirnov statistic~\cite{massey1951kolmogorov} is used to do the goodness-of-fit test (i.e. how close the transformed copula data is to the reference uniform distribution)~\cite{lilliefors1967kolmogorov}.

The high p-values in Table \ref{tab:table3} indicate that the copula data generated from PIT is uniformly distributed.
 % include figure, this is a floating figure

 \subsection{\label{sec:dependence} Measure of Dependence}
 \begin{figure*}[t]
\begin{center}
  \includegraphics[width = 10cm]{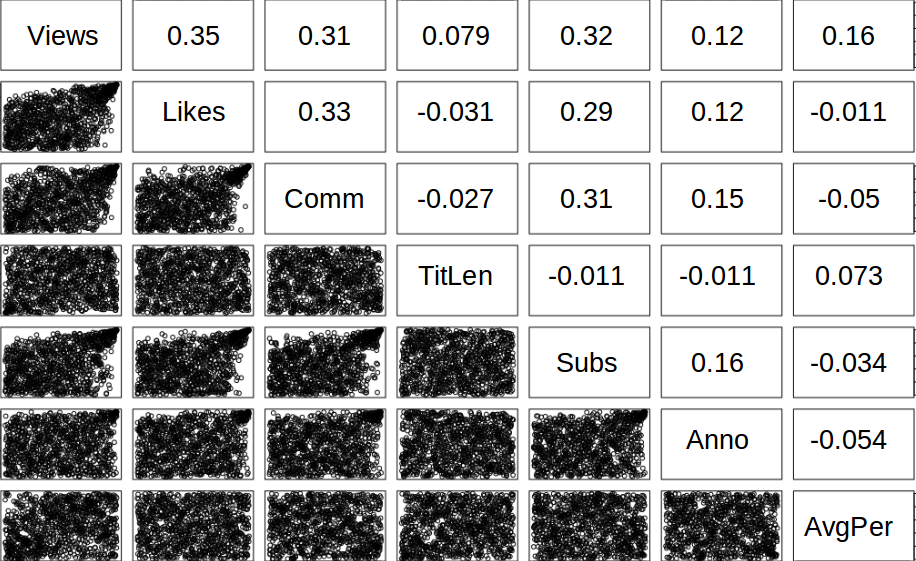}
\caption{\label{fig:scatter}{The scatter plots (lower matrix) and empirical Kendall's tau (upper matrix) for the copula data of \enquote{News} YouTube videos. The empirical Kendall's tau is estimated from data using Table II, while the scatter plots provide qualitative information regarding the pair-wise dependence. For instance, plots of \enquote{titlen} and \enquote{AvgPer} with other metrics are scattered everywhere, this implies weak dependence of these two metrics on the other metrics. On the other hand, scatter plots of \enquote{Views}, \enquote{Likes} and \enquote{Subs} show distinct patterns of co-movements with each other, this implies their strong dependence on each other, and might be in central position in the dependence structure.}}
\end{center}
\end{figure*}

For each pair-copula, instead of the classical correlation coefficient, the Kendall rank correlation coefficient (all called Kendall's tau)~\cite{Kendall1948rank} is a useful measure of dependence. The classical correlation coefficient for two random variables $X,Y$: $\rho_{X,Y} = \frac{E[(X-\mu_X)(Y-\mu_Y)]}{\sigma_X\sigma_Y}$ only detects linear dependencies, and it is not preserved by copulas. Therefore is not considered. Unlike the classical correlation coefficient, the Kendall's tau\footnote{Kendall's tau is a measure of rank correlation: the similarity of the orderings of the data when ranked by each of the quantities. The rank is high when the variables have a similar rank, and is low when the ranks are dissimilar.}
is preserved by copulas~\cite{venter2002tails}.

Given a dataset with $N$ observations and seven meta-level metrics, there are $\binom{7}{2}=21$ pairs of metrics in total; an estimate of dependency for $k^{th}$ pair of metrics $(X, Y)$ is obtained by the empirical Kendall's tau coefficient:
\begin{equation}
	\tau_{em}(X,Y)=\frac{A_N^k-B_N^k}{N(N-1)/2}.
\end{equation}
where $A_N^k$ and $B_N^k$ are the concordance and disconcordance number~\cite{genest2007everything} of $k^{th}$ pair that has $N$ data points. The concordance number and the disconcordance number count the data points that move in the same direction and opposite direction, respectively.

In contrast to the empirical Kendall's tau, the theoretical Kendall's tau, which is computed using bivariate copulas\cite{schweizer1981nonparametric}: 
\begin{equation}
\tau_{th}(X,Y) = 4\int_{u = 0}^{u = 1} \int_{v = 0}^{v = 1} C(u,v)dC(u,v)-1.
\end{equation}

Here, $C$ is the bivariate copula to be selected in the vine model, $u$ and $v$ represent the copula data of $X$ and $Y$, respectively.

The empirical Kendall's tau and scatter plot for the copula data of \enquote{News} videos are shown in Fig.\ref{fig:scatter}. In order to compute empirical Kendall's tau, 6000 \enquote{News} videos are randomly selected from 85657 videos over 3217 YouTube \enquote{News} channels, its theoretical Kendall's tau values shown in Table V(A) are computed from the vine copula model. The empirical Kendall's tau, which is computed from data, is required in the R-vine structure selection step for the pair copula construction of vine copula (reviewed in the Appendix); while theoretical Kendall's tau is computed as the measure of dependence given by the vine copula model.

\subsection{Summary of Main Conclusions}
% tif and gif file formats are NOT supported and must be converted to eps, jpg, or png

Based on the vine copula analysis of the YouTube dataset, we conclude several important facts regarding dependence structures of YouTube videos. (Sec.\ref{sec:results} describes these dependencies in more detail.)
\begin{enumerate}
\item Conditional dependence of the seven meta-level metrics of YouTube videos is insignificant, compared with a much stronger unconditional dependence.

\item Amongst the seven meta-level metrics of YouTube, \enquote{number of views} and \enquote{number of likes} are in the central position of the dependence structure. The other five metrics are much more dependent on these two metrics than on each other.

\item The \enquote{News} category has a different dependence structure from the other four categories. Furthermore, the metrics of \enquote{News} videos have stronger inter-dependence when their values are large (i.e. positive extreme co-movements), the other categories don't possess this property.

\item For \enquote{Gaming} video category, \enquote{annotation clicks rate} is least dependent on number of views among all the five categories.

\item For \enquote{Comedy} video category, users are more likely to subscribe to the channel for a popular video than other categories.
\item Users watching \enquote{Comedy} videos are more likely to give comments.

\item For a video with certain popularity, users are more likely to finish a video in \enquote{Fashion} category than in other categories.

\item \enquote{Length of video title} is independent of the other metrics.

\end{enumerate}

\subsection{\label{sec:results} Results and Discussion}
We are now ready to discuss the main conclusions on our dependency analysis of YouTube video meta-level metrics.

Once the empirical marginal distributions of YouTube video metrics are obtained using vine copula described in Sec.\ref{sec:marginal}, the three steps outlined in the Appendix were applied to model the multivariate distributions for YouTube video metrics. Table V summarizes the estimated parameters: Kendall's tau $\tau$, copula parameters and lower/upper tail dependence $\lambda_L$/$\lambda_U$ (defined in~(\ref{eqn:hey4}) in the Appendix), for the vine copula model of all five video categories. General dependence analysis for the five categories and category-specific conclusions are shown as follows.

\subsubsection{\label{sec:general_dependence} General Dependence Analysis for All Five Categories}
\begin{figure*}
\centering
\includegraphics[width=10cm]{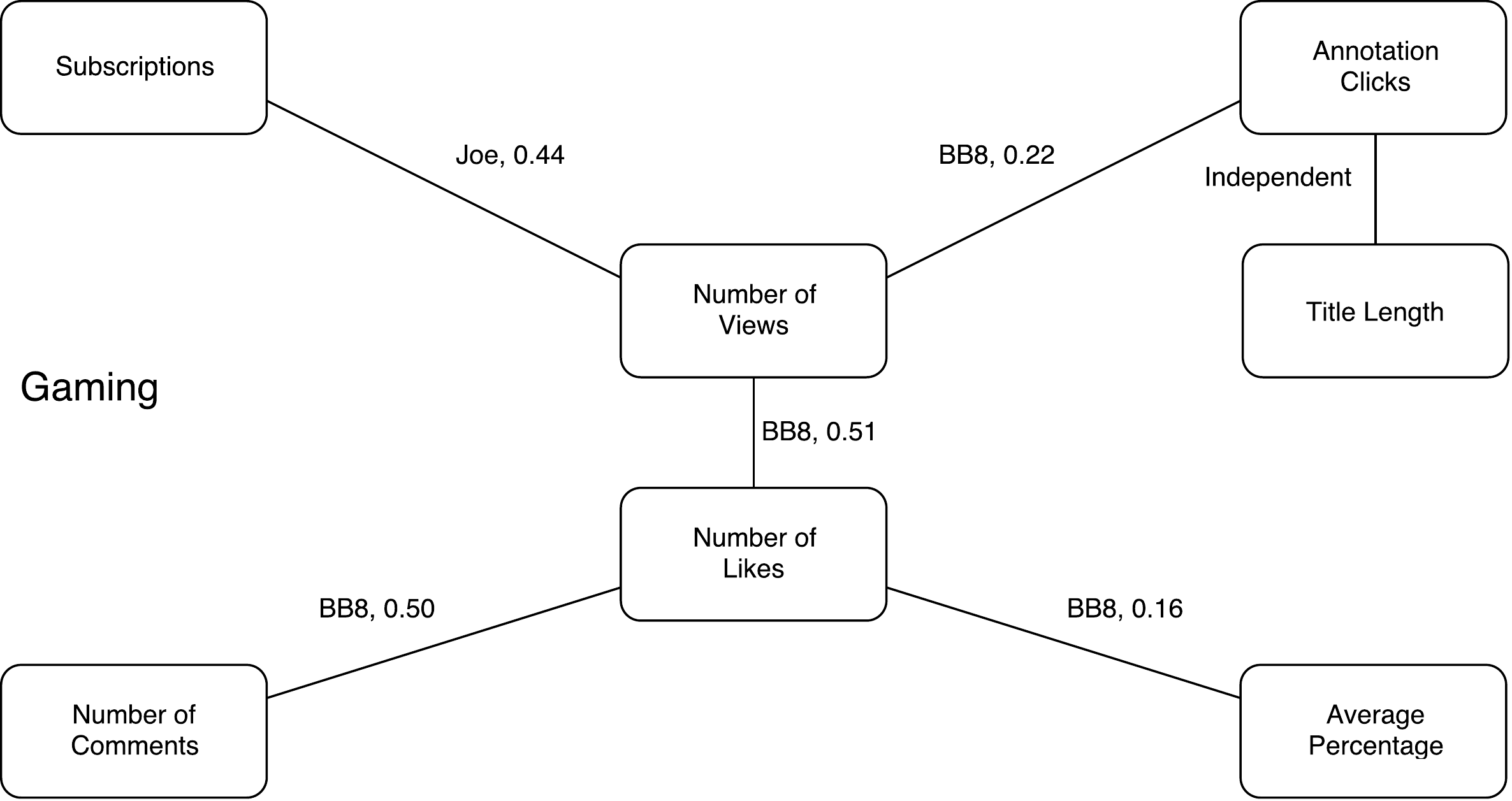}

\includegraphics[width=10cm]{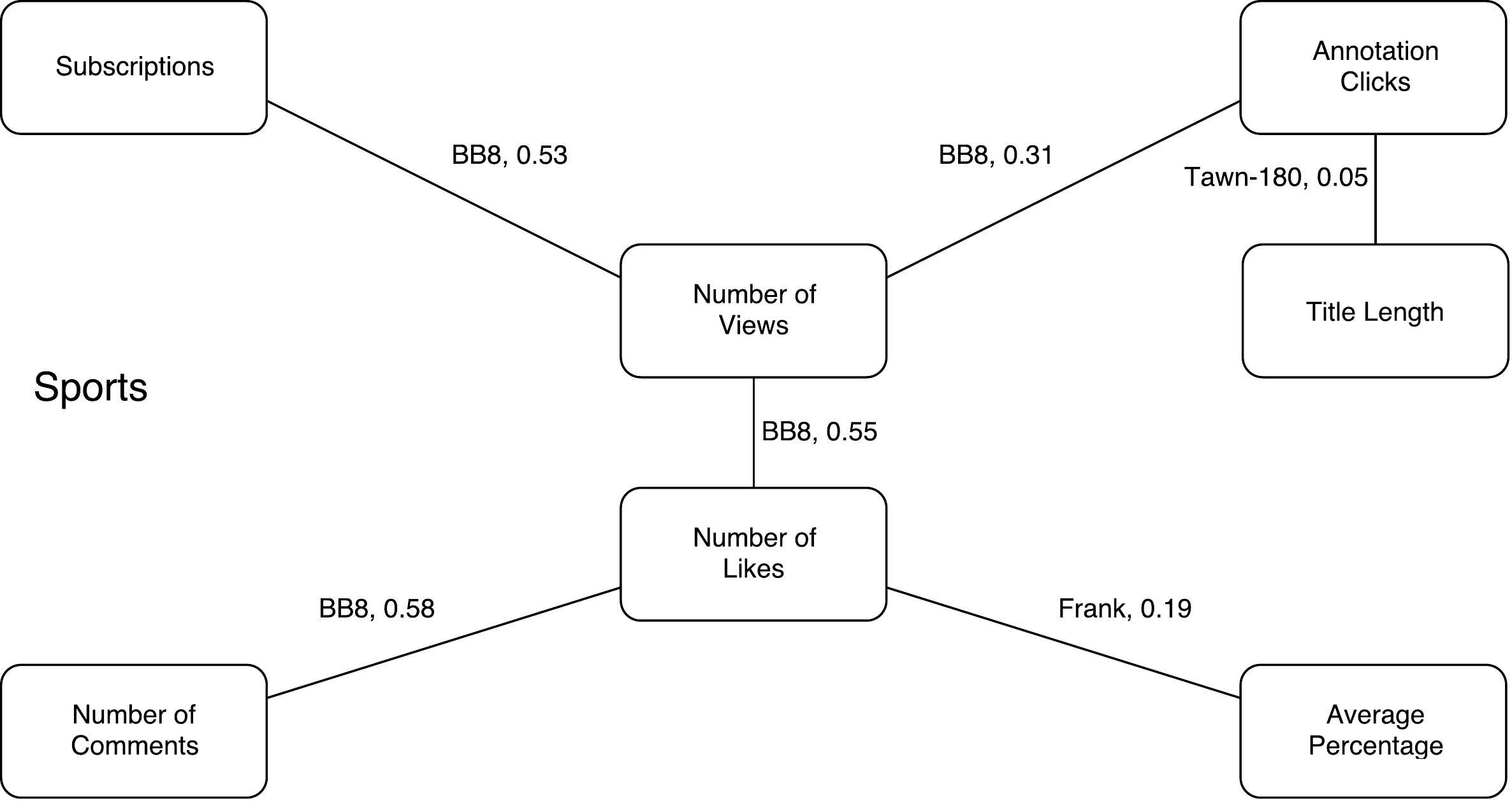}

\includegraphics[width=10cm]{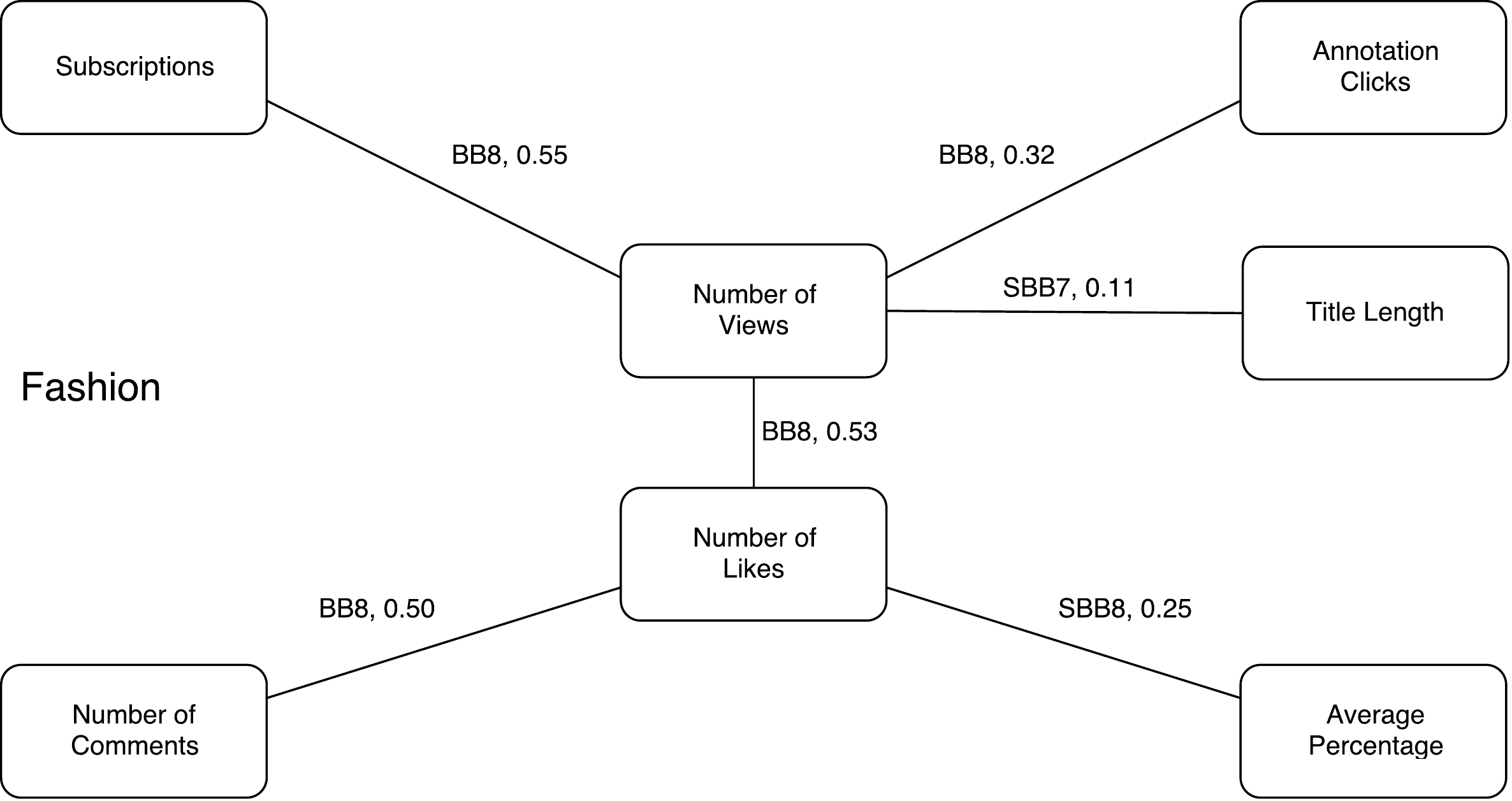}

\includegraphics[trim=0cm 3.3cm 0cm 3.3cm, clip=true, width=10cm]{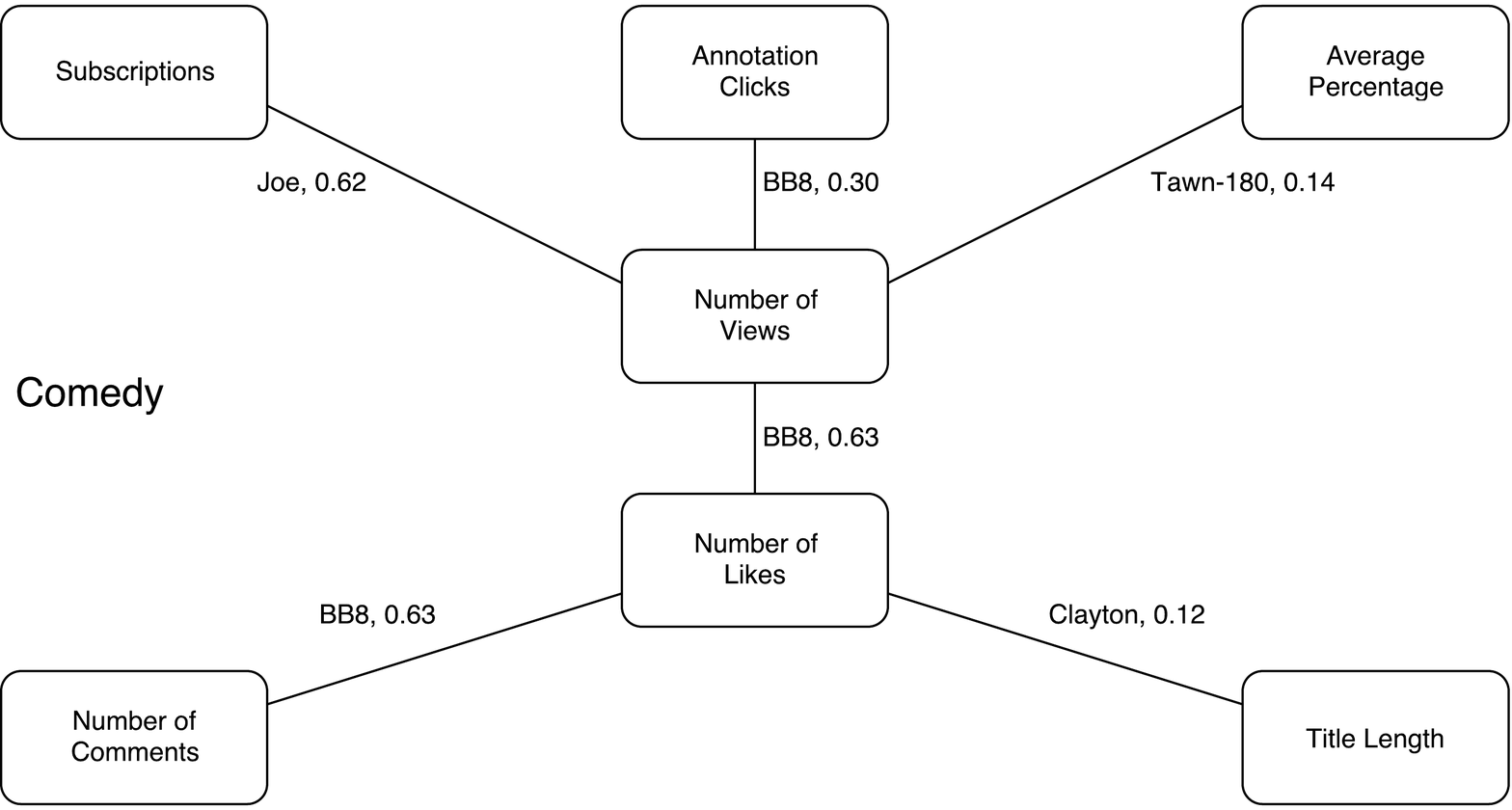}\quad
\vspace{0cm}

\caption{The first dependence (unconditional dependence) trees of the vine copula for the \enquote{Comedy}, \enquote{Sports}, \enquote{Fashion}, \enquote{Gaming} YouTube video data, along with their theoretical Kendall's tau values and pair-wise copula families. Second and further dependence trees are not provided, since we have shown in Table V that conditional dependence  is not significant in vine structure of YouTube videos.  It can be observed that \enquote{number of views} and \enquote{number of likes} are in the central position of the dependence structure.}
\label{fig:five}
\end{figure*}

Fig.\ref{fig:five} and Fig.\ref{fig:one} are the main graphical dependency results of our analysis.
They illustrate the dependence structure for all five video categories, with pair-wise theoretical Kendall's tau values and bivariate copula names.

First, as described in the Appendix, a vine structure uses multiple dependence trees to describe the dependence structure, which consists of two types of trees:
 \begin{compactitem}
\item Unconditional dependence tree: the first tree of the vine
\item Conditional dependence trees: all the trees of a vine excluding the unconditional dependence tree 
\end{compactitem} 
  However, in Fig.\ref{fig:five} and Fig.\ref{fig:one}, we only use the first trees (i.e. unconditional dependence trees) to represent the dependence structure for YouTube video metrics. Unconditional/conditional dependence trees describe the dependence of two or more random variables when another random variable is present, and are induced through the decomposition of vine structure. For instance, in~(\ref{eqn:hey5}), given three random variables $X_1$, $X_2$ and $X_3$, copula density $c_{12}$ is the joint density of ($X_1$, $X_2$), and $c_{13\vert 2}$ is the conditional density of $X_1$, $X_3$ given $X_2$. So $c_{12}$ describes the unconditional dependence of $X_1$, $X_2$, while $c_{13\vert 2}$ describes the conditional dependence of $X_1$, $X_3$ given $X_2$. Table V shows that the unconditional dependencies for the seven metrics (Kendall's tau values of 0.3-0.6) are much stronger than the conditional dependencies (Kendall's tau values of 0.01-0.2). Therefore, the conditional dependence is insignificant for YouTube video metrics, and can be ignored. 

 Second, in Fig.\ref{fig:five}, it can be observed that for \enquote{Comedy}, \enquote{Sports}, \enquote{Fashion}, \enquote{Gaming} video categories, \enquote{number of views} and \enquote{number of likes}	 are in the central position of the dependence structure of the seven YouTube video metrics. More specifically, \enquote{number of comments}, \enquote{number of subscribers} and \enquote{annotation clicks} have a strong unconditional dependence on these two video metrics, with high Kendall' tau values between 0.3 and 0.65. The exceptions are \enquote{title length} and \enquote{average percentage of watching}: \enquote{title length} shows little dependence on other metrics with Kendall's tau values that are no more than 0.1, while \enquote{Average Percentage of Watching} shows a moderate dependence on other metrics with Kendall's tau values between 0.14 and 0.25. 
 
Third, given the central roles of \enquote{number of views} and \enquote{number of likes} in the dependence structure, we should assign dominant weights to them if we want to assess a video using these seven video metrics. Note that \enquote{number of views} and \enquote{number of likes} are important indicators for the popularity of a video. Since \enquote{title length} is statistically independent of \enquote{number of views} and \enquote{number of likes} (considering its Kendall's tau values that are less than 0.1), one can conclude that changing video title length will not increase the popularity of video. As for \enquote{average percentage of watching}, which is an important indicator for video quality, surprisingly it is not highly dependent on the YouTube video viewcount.

\subsubsection{\label{sec:dependence_analysis} Category-specific Dependence Analysis}
\begin{figure*}
\begin{center}
  \vspace{0cm}\includegraphics[width = 10cm]{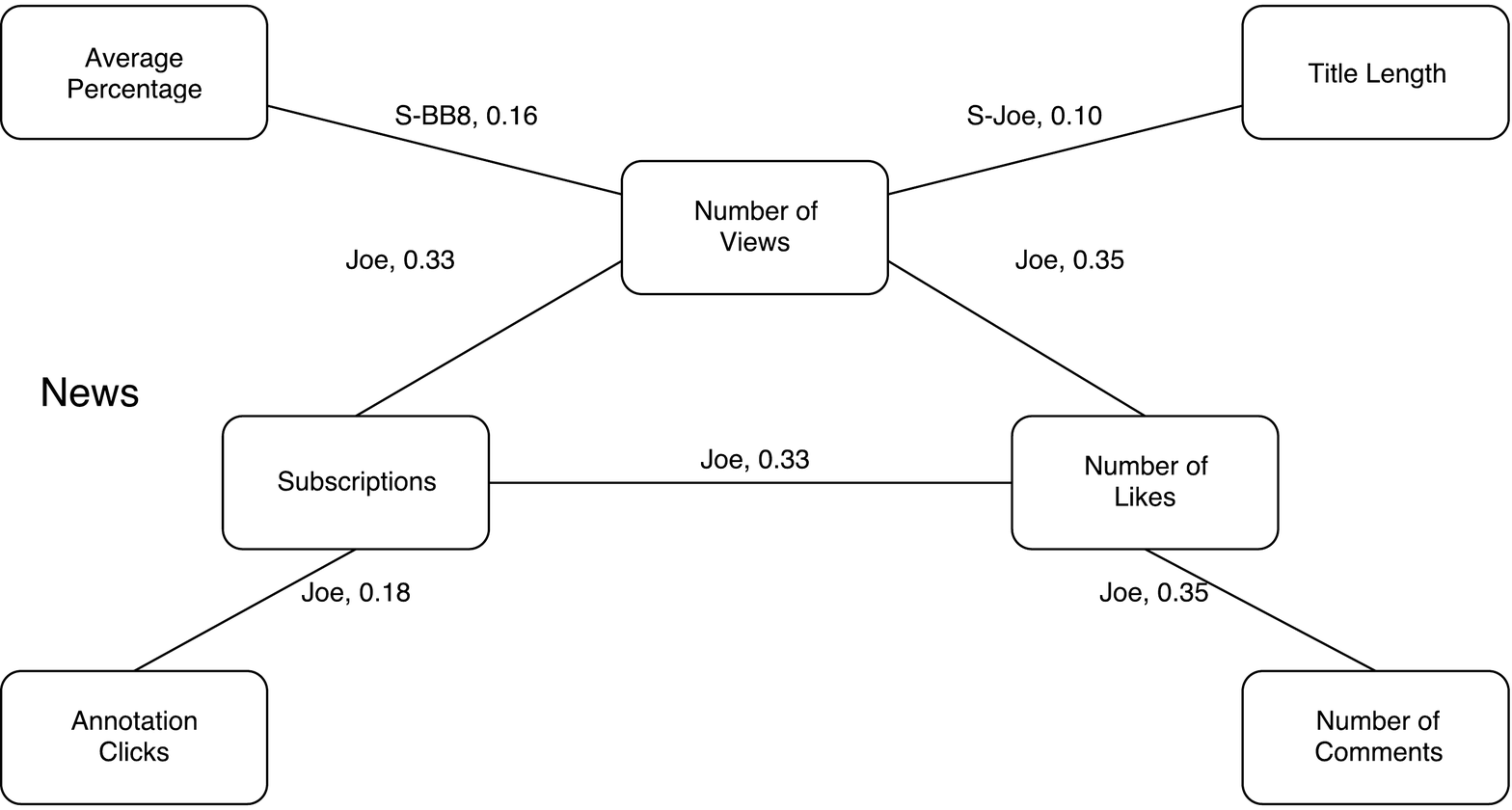}
\caption{\label{fig:one}{The first dependence (unconditional dependence) trees of the vine copula for the \enquote{news} category, with its theoretical Kendall's tau values and pair-wise copula families.  It can be observed that different from the other four categories,  in the dependence structure of the \enquote{News} category, \enquote{number of subscribers} is as important as \enquote{number of views} and \enquote{number of likes}. Although the overall inter-dependence of video metrics in \enquote{News} are weaker than in the other four categories.}}
\end{center}
\end{figure*}

By comparing the dependence structures of different categories in Fig.\ref{fig:five}, we can see that \enquote{Gaming}, \enquote{Sports}, \enquote{Fashion} and \enquote{Comedy} videos possess similar dependence structure. That is, \enquote{number of views} and \enquote{number of likes} are in the central position of the dependence structure, while other metrics are in the border of the dependence structure. The exception is the \enquote{News} category. As shown in Fig.\ref{fig:one}, \enquote{number of subscribers} for the \enquote{News} category also plays the same central role as \enquote{number of likes} and \enquote{number of views}, although the overall dependence of \enquote{News} category turns to be weaker (i.e. smaller Kendall's tau values) than the other categories. This can be explained by the fact that, all the other four categories possess a strong characteristic of entertainment, which \enquote{News} doesn't have much, thus a big difference of dependence structure should be expected.

Furthermore, by comparing the theoretical Kendall's tau values among different video categories, more category-specific conclusions can be made. The \enquote{Gaming} video category has the smallest Kendall's tau value (0.22) for the (Anno, Views) pair, while \enquote{Comedy} category has the largest $\tau$ for both the (Comm, Likes) pair and the (Subs,Views) pair. The largest $\tau$ value for the (AvgPer, Likes) pair belongs to the \enquote{Fashion} category. Therefore, YouTube viewers watching \enquote{Gaming} videos are more likely to accept the suggestions from the channel and watch more related videos; while \enquote{Comedy} channels should be a better place for video makers raising more subscribers and create active online community; if business looks for channels to put advertisements on, \enquote{Fashion} channels should have a better chance to make viewers look at their break-in advertisements. 

Finally, from Table V, it can be observed that for unconditional dependence structures of video categories including \enquote{Comedy}, \enquote{Sports}, \enquote{Fashion}, \enquote{Gaming}, there are few values for lower and upper tail dependence. This implies that for these categories, YouTube video metrics don't have many co-movements (i.e. small dependence) when their values are very large or very small. The exception is again the \enquote{News} category: with large values of upper tail dependence for four pairs of metrics shown in Table V(A), the \enquote{News} video metrics are expected to have stronger inter-dependence when the values of these metrics are large.

 \begin{table*} \centering
\caption{\label{tab:table3}p-values of uniform K-S test for YouTube data}
%\begin{ruledtabular}
\begin{tabular}{cccccccc}
Category & Views  & Likes & Comm & TitLen & Subs & Anno & AvgPer	\\ 
\hline % inserts single horizontal line
News & 0.79 & 0.97 & 0.72 & 0.91 & 0.98 & 0.86 & 0.91\\
\hline
Sports & 0.93 & 0.82 & 0.92 & 0.71 & 0.96 & 0.86 & 0.87\\
\hline
Fashion & 0.83 & 0.99 & 0.76 & 0.99 & 0.87 & 0.88 & 0.85\\
\hline
Gaming & 0.76 & 0.96 & 0.87 & 0.88 & 0.99 & 0.70 & 0.90\\
\hline
Comedy & 0.93 & 0.98 & 0.98 & 0.89 & 0.92 & 0.98 & 0.78\\
\end{tabular}
%\end{ruledtabular}
\end{table*}

\begin{table*}
 \label{tab:all_parameters}
  \caption{Estimated Parameters for the Vine Copula Model: bivariate copula types discussed in the Appendix, theoretical Kendall's tau $\tau$, bivariate copula parameters, lower/upper tail dependence $\lambda_L$/$\lambda_U$ (defined in~(\ref{eqn:hey4}) in the Appendix)}
  \begin{subtable}[h]{1\textwidth}
  \caption*{(A): News}
  \label{tab:parameter_news}
  
%    \begin{ruledtabular}
\begin{tabular}{cccccccc}
Tree  &Dependence Pair & Copula & $\tau$  & Parameters & $\lambda_L$ & $\lambda_U$	\\
\hline % inserts single horizontal line
1 & (Views, Comm) & S-Joe &0.098 &[1.20, null] &0.22 &0\\ % inserting body of the table
\hline
1 & (Likes, Comm) & Joe & 0.35 & [1.98, null] &0 &0.58\\
\hline
1 & (Views, Likes) & Joe & 0.35 & [1.95, null] &0 &0.57\\
\hline
1 & (Subs, Anno) & Joe & 0.18 & [1.42, null]&0 &0.37\\
\hline
1 & (Views, Subs) & Joe & 0.33 & [1.85, null] &0 &0.55\\
\hline
1 & (Views, Avgper) & S-BB8 & 0.16 & [1.77, 0.80] &0 &0\\
\hline
2 & (Titlen, Likes$|$Views) & BB8-270 & -0.081 & [-1.27, -0.91] &0 &0.55\\
\hline
2 & (Comm, Views$|$Likes) & S-BB7 & 0.11 & [1.10, 0.16] &0.12 &0.013\\
\hline
2 & (Likes, Subs$|$views) & Joe & 0.17 & [1.35, null] &0 &0.33\\
\hline
2 & (Anno, Views$|$subs) & Tawn-180 & 0.017 & [1.28, 0.020] &0.015 &0\\
\hline
2 & (Subs, Avgper$|$views) & BB8-90 & -0.121 & [-1.72, -0.72] &0 &0\\
\end{tabular}
%\end{ruledtabular}
  \end{subtable}

\begin{subtable}[h]{1\textwidth}
  \centering
  \caption*{(B): Comedy}
  \label{tab:parameters_comedy}
  
%\begin{ruledtabular}
\begin{tabular}{cccccccc}
Tree  &Dependence Pair & Copula & $\tau$  & Parameters & $\lambda_L$ & $\lambda_U$	\\
\hline % inserts single horizontal line
1 & (AvgPer, Views) & Tawn &0.14 &[1.78, 0.20] &0.17 &0\\ % inserting body of the table
\hline
1 & (Titlen, Likes) & Clayton & 0.12 & [0.26, null] &0.07 &0\\
\hline
1 & (Comm, Likes) & BB8 & 0.63 & [4.40, 0.98] &0 &0\\
\hline
1 & (Views, Views) & BB8 & 0.63 & [5.16, 0.91]&0 &0\\
\hline
1 & (Subs, Views) & BB8 & 0.62 & [4.53, 0.96] &0 &0\\
\hline
1 & (Likes, Anno) & BB8 & 0.30 & [1.95, 0.96] &0 &0\\
\hline
2 & (AvgPer, Comm$|$Views) & Frank & -0.17 & [-1.58, null] &0 &0\\
\hline
2 & (Titlen, Views$|$Likes) & Frank & -0.041 & [-0.37, null] &0 &0\\
\hline
2 & (Comm, Likes$|$Views) & Tawn & 0.12 & [1.30, 0.33] &0 &0.15\\
\hline
2 & (Views, Subs$|$Views) & Clayton & 0.13 & [0.3, null] &0 &0.10\\
\hline
2 & (Subs, Anno$|$Views) & BB8 & 0.0075 & [1.3, 0.84] &0 &0\\
\end{tabular}
%\end{ruledtabular}
\end{subtable}

\begin{subtable}[h]{1\textwidth}
  \centering
  \caption*{(C): Sports}
  \label{tab:parameters_sports}
  
%  \begin{ruledtabular}
\begin{tabular}{cccccccc}
Tree  &Dependence Pair & Copula & $\tau$  & Parameters & $\lambda_L$ & $\lambda_U$	\\
\hline % inserts single horizontal line
1 & (AvgPer, Likes) & Frank &0.19 &[1.80, null] &0 &0\\ % inserting body of the table
\hline
1 & (Comm, Likes) & BB8 & 0.58 & [3.94, 0.96] &0 &0\\
\hline
1 & (Likes, Views) & BB8 & 0.55 & [4.19, 0.90] &0 &0\\
\hline
1 & (Subs, Views) & BB8 & 0.53 & [3.47, 0.95]&0 &0\\
\hline
1 & (Views, Anno) & BB8 & 0.31 & [2.65, 0.79] &0 &0\\
\hline
1 & (Titlen, Anno) & Tawn-180 & 0.045 & [1.29, 0.099] &0.06 &0\\
\hline
2 & (AvgPer, Comm$|$Likes) & Gaussian & -0.052 & [-0.082, null] &0 &0\\
\hline
2 & (Comm, Views$|$Likes) & Tawn & 0.035 & [1.11, 0.22] &0 &0.53\\
\hline
2 & (Likes, Subs$|$Views) & Tawn & 0.17 & [1.38, 0.54] &0 &0.25\\
\hline
2 & (Subs, Anno$|$Views) & BB8 & 0.12 & [1.46, 0.88] &0 &0\\
\hline
2 & (Views, Titlen$|$Anno) & Student t & 0.0036 & [0.0051, 9.92] &0.0073 &0.0073\\
\end{tabular}
%\end{ruledtabular}
\end{subtable}
\end{table*}

\begin{table*}
\begin{subtable}[h]{1\textwidth}
  \centering
  \caption*{(D): Fashion}
  \label{tab:parameters_fashion}
  
%  \begin{ruledtabular}
\begin{tabular}{cccccccc}
Tree  &Dependence Pair & Copula & $\tau$  & Parameters & $\lambda_L$ & $\lambda_U$	\\
\hline % inserts single horizontal line
1 & (Likes, Comm) & BB8 &0.50 &[3.25, 0.95] &0 &0\\ % inserting body of the table
\hline
1 & (Views, Titlen) & Survival BB7 & 0.11 & [1.13, 0.12] &0.15 &0.003\\
\hline
1 & (Likes, AvgPer) & Survival BB8 & 0.25 & [2.06, 0.85] &0 &0\\
\hline
1 & (Views, Likes) & BB8 & 0.53 & [3.25, 0.97]&0 &0\\
\hline
1 & (Views, Subs) & BB8 & 0.55 & [3.46, 0.95] &0 &0\\
\hline
1 & (Views, Anno) & BB8 & 0.32 & [2.68, 0.81] &0 &0\\
\hline
2 & (Comm, Views$|$Likes) & Tawn & 0.10 & [1.25, 0.30] &0 &0.13\\
\hline
2 & (Titlen, Likes$|$Views) & BB8 & -0.053 & [-1.22, -0.81] &0 &0\\
\hline
2 & (AvgPer, Views$|$Likes) & Tawn-90 & -0.029 & [-1.17, 0.093] &0 &0\\
\hline
2 & (Likes, Subs$|$Views) & BB8 & 0.27 & [3.12, 0.67] &0 &0\\
\hline
2 & (Subs, Anno$|$Views) & Tawn & 0.068 & [1.17, 0.33] &0 &0.1\\
\end{tabular}
%\end{ruledtabular}
\end{subtable}

\begin{subtable}[h]{1\textwidth}
  \centering
  \caption*{(E): Gaming}
  \label{tab:parameters_gaming}
  
%    \begin{ruledtabular}
\begin{tabular}{cccccccc}
Tree  &Dependence Pair & Copula & $\tau$  & Parameters & $\lambda_L$ & $\lambda_U$	\\
\hline % inserts single horizontal line
1 & (AvgPer, Likes) & BB8 &0.16 &[2.22, 0.65] &0 &0\\ % inserting body of the table
\hline
1 & (Comm, Likes) & Joe & 0.50 & [2.82, null] &0 &0.72\\
\hline
1 & (Likes, Views) & BB8 & 0.51 & [2.96, 0.99] &0 &0\\
\hline
1 & (Subs, Views) & Joe & 0.44 & [2.42, null]&0 &0.67\\
\hline
1 & (Views, Anno) & BB8 & 0.22 & [1.59, 0.97] &0 &0\\
\hline
1 & (Titlen, Anno) & Gaussian & 0.037 & [0.06, null] &0 &0\\
\hline
2 & (AvgPer, Views$|$Likes) & Student t & -0.034 & [-0.054, 7.19] &0.016 &0.016\\
\hline
2 & (Comm, Views$|$Likes) & Student t & 0.13 & [0.21, 16.5] &0.003 &0.003\\
\hline
2 & (Likes, Subs$|$Views) & Gumbel & 0.17 & [1.21, null] &0 &0.23\\
\hline
2 & (Subs, Anno$|$Views) & Frank & 0.046 & [0.41, null] &0 &0\\
\hline
2 & (Views, Titlen$|$Anno) & Tawn-180 & 0.014 & [1.11, 0.045] &0.016 &0\\
\end{tabular}
%\end{ruledtabular}
\end{subtable}

\end{table*}

\begin{table}
\caption{\label{tab:simulation}p-values of White test for YouTube copula data}
%\begin{ruledtabular}
\begin{tabular}{cccccccc}
Categories &News&   Sports&   Fashion& Gaming   & Comedy   	\\
\hline
p-values & 0.674 & 0.685 & 0.591 & 0.714 & 0.666  \\
\end{tabular}
%\end{ruledtabular}
\end{table}
%\end{space}

\begin{figure*}[t]
\begin{center}
  \includegraphics[width = 10cm]{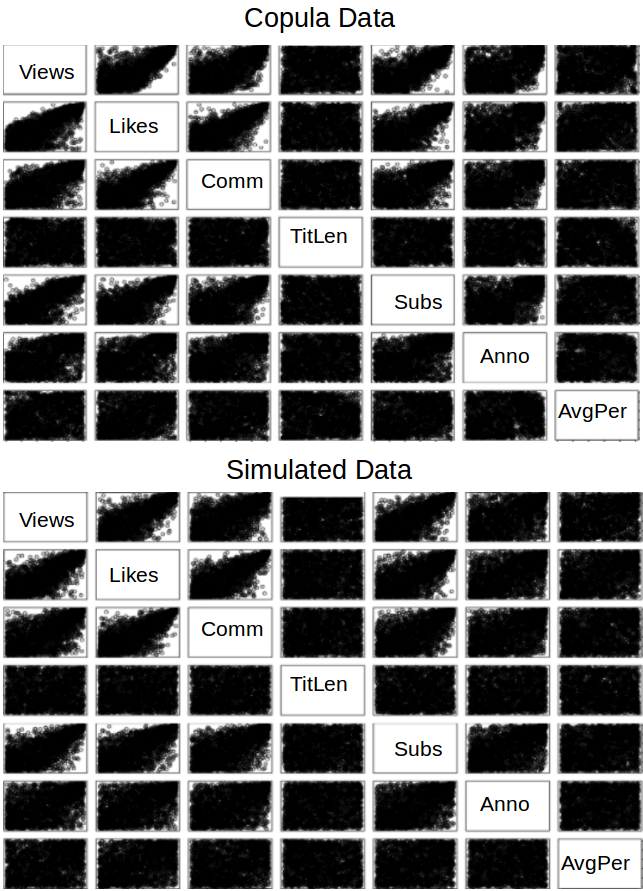}
\caption{\label{fig:simulation_compare}{The scatter plots for original and simulated \enquote{Sports} YouTube copula data. It can be observed that, the original YouTube copula data and the copula data simulated from vine copula model share a dependence pattern. This qualitatively shows that the vine copula model captures the dependence patterns accurately, such conclusion is validated by the high p-values of White test in Table \ref{tab:simulation}.}}
\end{center}
\end{figure*}

\section{\label{sec:simulation}  Validation of Vine Copula Model Using the White Test}
 How good is the vine copula model of Sec.\ref{sec:analysis} in terms of determining the dependency structures in YouTube? Goodness-of-fit tests including the misspecification test, the information matrix ratio test, and the Rosenbatt's transform test, have been introduced for the vine copulas over the last decade~\cite{genest2009goodness}. Considering that we used regular vine (R-vine) to model the dependence structure, below we use one type of misspecification test (we call it \enquote{White test}) for the goodness-of-fit test of R-vine. Proposed by White and enhanced by Schepsmeier, this test is shown to have excellent performance and power behavior~\cite{schepsmeier2013goodness}.
 
 The theorem of White states that, under the correct model, the negative Hessian matrix and outer product of gradient for the likelihood function are equal~\cite{schepsmeier2013goodness}. Therefore, the White test is the following hypothesis test:
 \begin{equation*}
 H_0: \mathbb{H}(\boldsymbol{\theta_0})+\mathbb{C}(\boldsymbol{\theta_0})=0
 \end{equation*}
\begin{equation*}
H_1: \mathbb{H}(\boldsymbol{\theta_0})+\mathbb{C}(\boldsymbol{\theta_0}) \neq 0.
\end{equation*}
 where $\mathbb{H}(\boldsymbol{\theta})$ is the Hessian matrix:
\begin{equation}
\mathbb{H}(\boldsymbol{\theta}) = \frac{\partial ^2}{\partial^2 \boldsymbol{\theta}}L(\boldsymbol{\theta} \vert \boldsymbol{U}).
\end{equation}  
   and $\mathbb{C}(\boldsymbol{\theta})$ is the outer product of the gradient:
 \begin{equation}
 \mathbb{C}(\boldsymbol{\theta}) = \frac{\partial}{\partial \boldsymbol{\theta}}L(\boldsymbol{\theta} \vert \boldsymbol{U})(\frac{\partial}{\partial \boldsymbol{\theta}}L(\boldsymbol{\theta}\vert \boldsymbol{U}))^T.
 \end{equation}
$L(\boldsymbol{\theta} \vert \boldsymbol{U})$ is the likelihood function defined in (\ref{eqn:MLE}), and $\boldsymbol{U}$ is the simulated copula data following our vine copula model. The p-values of the White test can be computed based on (4) and (5), the reader can refer to~\cite{schepsmeier2013goodness} for more details.

Table \ref{tab:simulation} lists the p-values of the White test for the vine copula model, given the YouTube dataset. Based on the high p-values of White test for all five YouTube video categories considered, the null hypothesis can not be rejected. Thus the vine copula model obtained from Sec.\ref{sec:results} is a reasonable model for the multivariate probability distribution of YouTube video metrics. Fig.\ref{fig:simulation_compare} compares the pairwise scatter plots of YouTube copula data, along with the simulated data sampled from our vine copula model. A common dependence structure pattern can be observed in both plots, which show that the vine copula model captures the dependence structures of YouTube data accurately.

\section{YouTube Dynamics: Causality and Scheduling.}
\label{sec:causal:relation:YouTube}
Thus far we have unravelled dependency structures in YouTube metrics using vine copulas. Recall that Fig.\ref{fig:five} shows the dependence
between view count and number of subscribers.
In this section we dig further into the dynamics of YouTube in two ways: 
First we use Granger causality (see also \cite{hoiles2017engagement} for a more detailed study)  to detect the causal relationship between subscriber and viewer counts and how it can be used to estimate the next day subscriber count of a channel. The results are important  for determining the popularity of a YouTube channel. Second, we study the scheduling dynamics of  of YouTube channels. 
We find the interesting property that for popular gaming YouTube channels with a dominant upload schedule, deviating from the schedule increases the views and the comment counts of the channel.

\subsection{Causality Between Subscribers and View Count in YouTube}

\begin{figure}[t]
	\centering
	\includegraphics[scale=0.8]{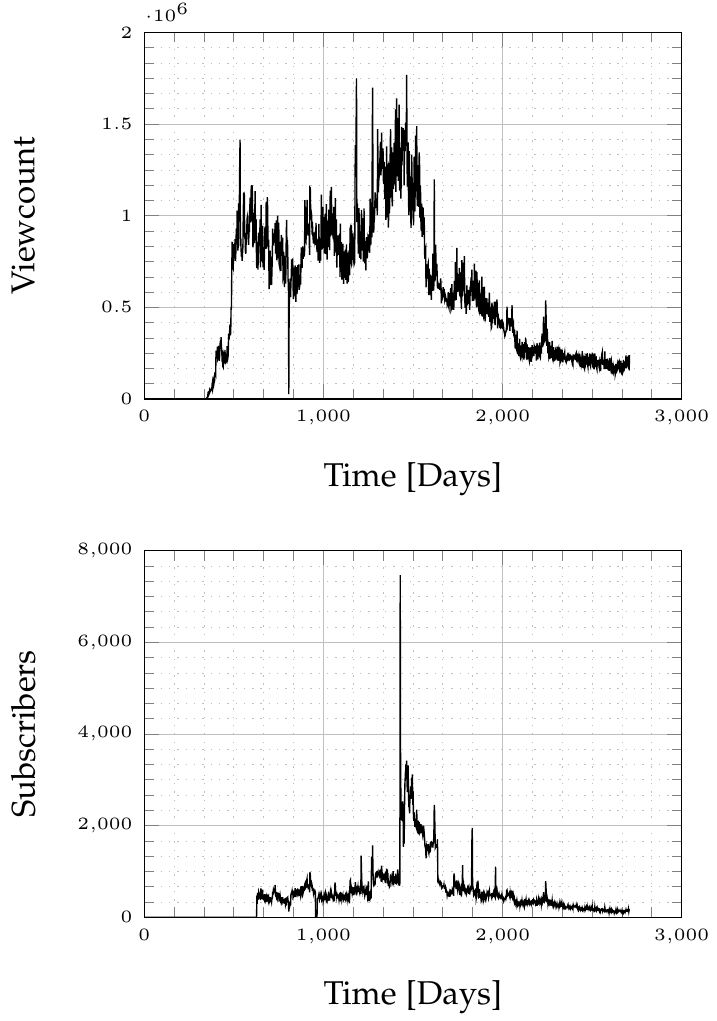}
	\caption{Viewcount and subscribers for the popular movie trailer channel: VISOTrailers. The Granger causality test for view counts ``Granger causes'' subscriber count is true with a p-value of $5\times 10^{-8}$. }
	\label{fig:Channel6:sub:view}
\end{figure}

Fig.~\ref{fig:Channel6:sub:view} displays the subscriber and view count dynamics of a popular movie trailer channel in YouTube. 
It is clear from Fig.~\ref{fig:Channel6:sub:view} that the subscribers ``spike'' with a corresponding ``spike'' in the view count. 

The main idea behind Granger causality is that if the value(s) of a lagged time-series can be used to predict another time-series, then the lagged time-series  ``Granger cause'' the predicted time-series. 
%In the YouTube social network it was found that a first-order model can sufficiently be used to establish the Granger causality relationship between previous subscriber and viewer counts to predict the next day subscriber count. 
%
To formalize the Granger causality model, let $s^j(t)$ denote the number of subscribers to a channel $j$ on day $t$, and $v_i^j(t)$ the corresponding view count for a video $i$ on channel $j$ on day $t$. The total number of videos in a channel on day $t$ is denoted by $\mathcal{I}(t)$. Define, 
\begin{equation}
	\hat{v}^j(t) = \sum_{i=1}^{\mathcal{I}(t)}v_i^j(t),
\end{equation}
as the total view count of channel $j$ at time $t$. 
The Granger causality test involves testing if the coefficients $b_i$ are non-zero in the following equation which models the relationship between subscribers and view counts: 
\begin{equation}
	s^j(t) = \sum_{k = 1}^{n_s}a_k^j s^j(t-k) + \sum_{i=k}^{n_v}b_k^j \hat{v}^j(t-k) + \varepsilon^j(t),
	\label{eqn:arma:model}
\end{equation}
where $\varepsilon^j(t)$ represents normal white noise for channel $j$ at time $t$. The parameters $\{a_i^j\}_{\{i=1,\dots,n_s\}}$ and $\{b_i^j\}_{\{i=1,\dots,n_v\}}$ are the coefficients of the AR model in~\eqref{eqn:arma:model} for channel $j$, with $n_s$ and $n_v$ denoting the lags for the subscriber and view counts time series respectively. 
If the time-series $\mathcal{D}^j=\{s^j(t),\hat{v}^j(t)\}_{t\in\{1,\dots,T\}}$ of a channel $j$ fits the model~\eqref{eqn:arma:model}, then we can test for a causal relationship between subscribers and view count. 
In equation (\ref{eqn:arma:model}), it is assumed that $|a_i| < 1$, $|b_i| < 1$ for stationarity. The causal relationship can be formulated as a hypothesis testing problem as follows:  
\begin{equation}
	%H_0: b_1\neq 0 \text{ and } a_1\neq 0 \text{ vs. } H_1: b_1= 0 \text{ and } a_1\neq 0. 
	H_0: b_1 = \dots = b_{n_v}=0 \text{ vs. } H_1: \text{Atleast one } b_i \neq 0. 
	\label{eqn:hypothesis}
\end{equation}
The rejection of the null hypothesis, $H_0$, implies that there is a causal relationship between subscriber and view counts.

Two hypothesis testing procedures are applied to test for Granger causality. 
First, we use Box-Ljung test  is to evaluate the quality of the model~\eqref{eqn:arma:model} for the given dataset $\mathcal{D}^j$. If satisfied, then the Granger causality hypothesis~(\ref{eqn:hypothesis}) is evaluated using the Wald test. If both hypothesis tests pass then we can conclude that the time series $\mathcal{D}^j$ satisfies Granger causality--that is, the previous day subscriber and view count have a causal relationship with the current subscriber count. 

An important issue prior to performing the Granger causality test is to determine the percentage of videos in the YouTube dataset  satisfy the AR model in~(\ref{eqn:arma:model}). To perform this analysis we apply the Box-Ljung test with a confidence of 0.95 (p-value = $0.05$). 
%For this confidence, we found that $80\%$ of the channels subscriber dynamics follow the AR(1) model (\ref{eqn:arma:model}). Insight into the number of channels that satisfy the AR(1) model for different confidence values are provided in Fig.~\ref{X}. As seen from Fig.~\ref{X}, for $95\%$ of the channel subscribers to satisfy an AR(1) model we would require a confidence level of 0.8. The results in Fig.~\ref{X} illustrate the efficacy of the AR(1) model (\ref{eqn:arma:model}) for describing the subscriber dynamics. 
%We varied $n_s$ and $n_v$, the number of lags for subscribers and view counts respectively, and found that for $n_s = n_v = 3$, $91\%$ of the channels follow the model in~\eqref{eqn:arma:model}. 
First, we need to choose  $n_s$ and $n_v$, namely, the number of lags for the subscribers and view count time series. For $n_s=n_v = 1$, we found that only $20\%$ of the channels satisfy the model~(\ref{eqn:arma:model}). When $n_s$ and $n_v$ are increased to $2$, the number of channels satisfying the model increases to $63\%$.  For $n_s=n_v = 3$, we found that $91\%$ of the channels satisfy the model~(\ref{eqn:arma:model}), with a confidence of $0.95$ (p-value = $0.05$). Hence,  we chose $n_s=n_v = 3$. It is interesting to note that the mean value of coefficients $b_i$ decrease as $i$ increases indicating that older view counts have less influence on the subscriber count. Similar results also hold for the coefficients $a_i$. Hence, as expected, the previous day subscriber count and the previous day view count most influence the current subscriber count.  

Given the above causal relation between subscriber count and view count,
a natural question is:  Is there a causal relationship between the subscriber dynamics and the view count dynamics?
%(e.g. the hypothesis~\eqref{eqn:hypothesis}). 
This is modeled using the hypothesis in~(\ref{eqn:hypothesis}). 
We use the Wald test with a confidence of $0.95$ (p-value = $0.05$) and found that approximately $55\%$ of the channels satisfy the hypothesis. For approximately $55\%$ of the channels that satisfy the AR model (\ref{eqn:arma:model}), the view count ``Granger causes'' the current subscriber count. An interesting property is that the percentage of channels that satisfy Granger causality vary widely
depending on the channel category. For example, 67\% of Sports channels and 60\% of the Gaming  channels satisfy Granger causality, while  $80\%$ of Sports channels satisfy Granger causality. These results illustrate the importance of channel owners to not only maximize their subscriber count, but to also upload new videos or increase the views of old videos to increase their channels popularity (via increasing their subscriber count).

\subsection{Scheduling Dynamics}

From the dataset, we selected video channels with a \emph{dominant} upload schedule, as follows: The dominant upload schedule was identified by 
evaluating  the periodogram of the upload times of the channel and  comparing the highest value to the next highest value. 
If the ratio defined above is greater than $2$, we say that the channel has a dominant upload schedule. 
From the dataset containing $25$ thousand channels, only $6500$ channels contain a dominant upload schedule. 
Some channels, particularly those that contain high amounts of copied videos such as trailers, movie/TV snippets upload videos on a daily basis. 
These have been removed from the analysis  so as to  concentrate on those channels that contain only user generated content. 

We found that channels with gaming content account for $75\%$ of the $6500$ channels with a dominant upload schedule and the main tags associated with the videos were: ``game'', ``gameplay'' and ``videogame''.
We computed the average views when the channel goes off the schedule and found that on an average when the channel goes off schedule the channel gains views $97\%$ of the time and the channel gains comments $68\%$ of the time.  
Thus  channels with ``gameplay'' content with  periodic upload schedule  benefit from going off the schedule.  This suggests that by deliberately going
off schedule, gameplay channels can increase their view count.

\section{\label{sec:conclusions} Conclusions}
YouTube video metrics across thousands of channels  form a big-data time series.
In this paper, we conducted a data-driven dependence analysis for seven meta-level metrics of YouTube videos among five different categories, based on a YouTube dataset of over 6 million videos across 25 thousand channels.
To unravel the dependency structures of these meta-level metrics, we constructed a vine copula model on the YouTube dataset: the marginal distribution for each video metric is estimated empirically; the Kendall's tau is introduced as the measure of dependence in vine copula model; the vine structure is selected based on absolute empirical Kendall's tau, bivariate copulas are chosen by using AIC approach, then copula parameters are estimated using the method of maximum-likelihood. Parameters including theoretical Kendall's tau and tail dependence coefficients are computed for the dependence analysis of YouTube video metrics.

The analysis in this paper reveals three main conclusions regarding the YouTube users' watching behaviours in different video categories: conditional dependence is insignificant in YouTube video metrics, number of views and number of likes are in the central position of dependence structure, and \enquote{News} category possess a stronger tail dependence than the other YouTube video categories.

Finally, we also studied the dynamics of YouTube in two ways.  First, using Granger causality, we unravelled the causal dependence
between subscribers and view count. Second, by studying the upload dynamics, we found that periodic content providers can increase their view count
by going off schedule.

%\begin{acknowledgments}
%This research was supported by  ...
%\end{acknowledgments}

% --------------------------------------------------------------------------------------------------------------------------------

 % -------------------------------------------------------------------------------------------------------------------
 %   Appendix  (optional)

\appendix
\section{\label{sec:appendix}  Review of Vine Copula}

Recalling the definitions of copula and Sklar  theorem in Section~\ref{sec:shortreview}, this appendix gives a short outline
of vine copulas and their construction (which was used to analyze the YouTube dataset).

\subsection{Bivariate Archimedean Copula}
\label{sec:bi_archi}
One widely used copula type is the \enquote{Archimedean} copula:

\begin{def*}
Bivariate Archimedean Copula is a $[0,1]^2 \rightarrow [0,1]$ function $C$ with the form:
\begin{equation}
C(u,v) = \phi^{-1}(\phi(u)+\phi(v)).
\end{equation}
where $\phi(t)$ is called the generator function, which is a continuous, strictly decreasing function such that $\phi(1) = 0$; $\theta$ is the single parameter to be estimated.
\end{def*}
Bivariate archimedean copulas can be easily extended to arbitrarily high dimensions, but only bivariate ones are used for current copula applications, since it is just not a good idea that only one parameter is used to capture high dimensional dependence.

Archimedean copula includes several bivariate copulas such as Joe, Frank and Clayton copulas used in this work. Although there are over forty different types of bivariate copulas that are widely used. it is a harder problem to develop higher dimensional copulas. The “pair-copula construction” based on vine structure is a popular method of constructing a high dimensional copula~\cite{aas2009pair}.  
%
%Two more illustrative examples are Clayton copula and Joe copula used in this work; the clayton copula:
%\begin{equation}
%C_{\theta}^{CL}(u,v) = [u^{-\theta}+v^{-\theta}-1]^{-1/\theta}
%\end{equation}
%with a generator function of $\phi(t)=(1/\theta)(t^{-\theta}-1)$, appears to be a good dependence model for lower co-movements with little upper co-movements  of two metrics(i.e. asymmetric dependence); while the joe copula:

%\begin{align}
%C_{\theta}^{J}(u,v) = & 1-[(1-u)^{\theta}+(1-v)^{\theta} \\
%&-(1-u)^{\theta}(1-v)^{\theta}]^{1/\theta} \nonumber
%\end{align}

%with a generator function of $\phi(t)=-ln[1-(1-t)^{\theta}]$, is close to be the reverse of clayton copula, that is, it models two metrics with stronger upper co-movements with weak lower co-movements. Fig.\ref{fig:archi} shows the scatter plots of both clayton and joe copulas when $\theta =3$. 

%\begin{figure}
%  \centering
%      \includegraphics[origin=c, scale=0.25]{archi.jpg}
%  \caption{The scatter plots of Clayton copula(left) and Joe copula(right) when their single copula parameter $\theta = 3$. Clayton and Joe copula capture the lower and upper co-movements of two variables, respectively.}
%  \label{fig:archi}
%\end{figure}
%

\subsection{Pair Copula Construction}
\label{sec:pair_copula}
 For convenience, we illustrate the construction process for three random variables scenario, based on which a general $n$-variable case can be easily obtained. The density of a $3$-dimensional copula function $C$ is defined as:
\begin{equation}
c_{123}(u_1,u_2,u_3)=\frac{\partial C_{123}(u_1,u_2,u_3)}{\partial u_1 \partial u_2 \partial u_3 }.
\end{equation}
Then by~(\ref{eqn: hey1}), if we let $u_1=F_1(x_1),u_2=F_2(x_2),u_3=F_3(x_3)$, the density of joint distribution $F(x_1,x_2,x_3)$ can be represented as:
\begin{align}
\label{eqn:decomposition}
f(x_1,x_2,x_3)=&f_1(x_1)\cdot f_2(x_2) \cdot f_3(x_3) \\
&\cdot c_{123}(F_1(x_1),F_2(x_2),F_3(x_3)). \nonumber
\end{align}
One possible decomposition from the chain rule of probability density gives:
\begin{align}
\label{eqn:hey2}
f(x_1,x_2,x_3) = f_1(x_1)f_{2\vert 1}(x_2\vert x_1)f_{3\vert 12}(x_3\vert x_1,x_2).
\end{align}
Note that similar to~(\ref{eqn:decomposition}), the conditional probability densities can be represented as:
\begin{align}
\label{eqn:hey3}
f_{2\vert 1}(x_2 \vert x_1) &= c_{12}(F_1(x_1),F_2(x_2))f_2(x_2) \\
 f_{3\vert 2}(x_3 \vert x_2) &= c_{23}(F_2(x_2),F_3(x_3))f_3(x_3) \nonumber \\
f_{3\vert 12}(x_3\vert x_1, x_2) &= c_{13\vert 2}(F_{1\vert  2}(x_{1}\vert x_2),F_{3\vert 2}(x_3 \vert x_2)) \nonumber \\ 
& \cdot f_{3\vert 2}(x_3 \vert x_2). \nonumber
\end{align}
By combining~(\ref{eqn:hey2}) and~(\ref{eqn:hey3}), the multivariate probability density can be represented as multiplications of bivariate copula densities and marginal probability densities (recall the deifnitons of $c_{12}$ and $c_{13\vert 2}$ in Sec.~\ref{sec:general_dependence}):
\begin{align}
\label{eqn:hey5}
f(x_1,x_2,x_3) &= f_1(x_1)f_2(x_2)f_3(x_3) \\
& \cdot c_{12}(F_1(x_1),F_2(x_2)) \cdot c_{23}(F_2(x_2),F_3(x_3)) \nonumber \\
& \cdot c_{13\vert 2}(F_{1\vert  2}(x_{1}\vert x_2),F_{3\vert 2}(x_3 \vert x_2)). \nonumber
\end{align}

%The decomposition in~(\ref{eqn:hey5}) is only one of six possible decompositions. More generally, for an $n$-dimensional probability density, there are $n(n-1)(n-2)!2^{(n-2)(n-3)/2}/2$ possible decompositions in total~\cite{morales2010number}.

 In order to obtain a systematic way to represent these decompositions, Bedford and Cooke~\cite{bedford2001probability} proposed the so-called \enquote{Regular Vine} (or simply \enquote{R-Vine}) graphical structure, hence the name \enquote{vine-copula}. \enquote{Vine} refers to a nested set of tree structures, where the edges of the $i^{th}$ tree are the nodes of the ${i+1}^{th}$ tree. The formal definition of a regular vine structure is provided below.
\begin{defvine*}
A graphical structure $V$ is a regular vine of $m$ elements if:
\begin{enumerate}
\item $V = (T_1,...,T_{m-1})$ and all trees are connected
\item The first tree $T_1$ has node set $N_1 = {1,...,m}$ and edge set $E_1$; then for the next trees $T_i$, $i \in {2,...m-1}$, $T_i$ has the node set $N_i = E_{i-1}$
\item Proximity condition: If two nodes are connected by an edge in the ${i+1}^{th}$ tree, then the two edges in $i^{th}$ tree corresponding to these two nodes share a node.
\end{enumerate}

\end{defvine*}

\begin{figure}
\begin{center}
  \includegraphics[width = 8cm]{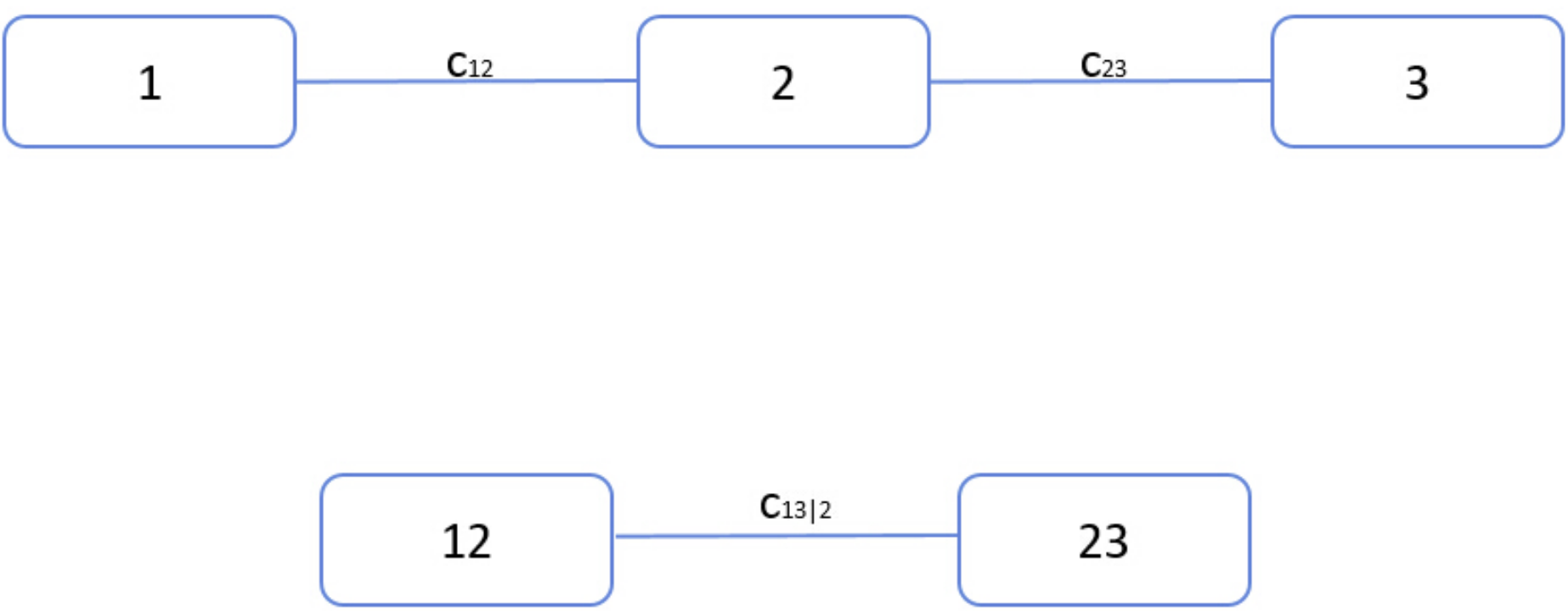}
\caption{\label{fig:bivariate}{The vine structure for a three-variable case. The vine corresponds to~(\ref{eqn:hey5}): the first tree represents the unconditional pairs $c_{12},c_{23}$, while the second tree represents the conditional pair $c_{13 \vert 2}$.}}
\end{center}
\end{figure}

In copula applications, the trees are used to represent the dependence structure of multiple variables: each edge represents the bivariate copula connecting one pair of marginals. For instance, the regular vine structure for three variables is shown in Fig.\ref{fig:bivariate}.

Once the uniformly distributed marginals are obtained, we can construct the vine copula model for a multivariate distribution.
Dissmann et al.~\cite{dissmann2013selecting} summarizes an inference procedure for pair-copula construction of vine copula. The following steps are required to construct a regular vine copula (R-vine) specification to a given dataset:

\begin{enumerate}
\item Construct the R-Vine structure by choosing an appropriate unconditional and conditional pair of metrics to use for vine copula model.

\item Choose a bivariate copula family for each pair selected in step (1).

\item Estimate the corresponding parameters for each bivariate copula.
\end{enumerate}

\subsubsection{R-Vine Structure Selection }

%In order to find the optimal R-vine structure for dependence modelling, one approach is to try all possible R-vine constructions and find the best fit for the given dataset; this is computationally intractable for high-dimensional problems, since the number of R-vines grows exponentially with the number of marginals $n$ (i.e. $n!/2\cdot 2^{n-2 \choose 2}$)~\cite{morales2011counting}. To resolve this issue,

 Dissman et al.~\cite{dissmann2013selecting} suggested a sequential tree-by-tree selection method, where the first tree of the regular vine is selected to be the one with strongest dependencies, then the following trees are selected based on the same criteria. Such a heuristic approach is not guaranteed to capture the global optimum point. However, since the first tree of R-vine (unconditional dependence tree) has the greatest influence on the model fit, the sequential method has a reasonable trade-off between efficiency and accuracy. A maximum spanning tree (MST) algorithm is proposed for R-vine structure selection, where the the spanning tree with the maximum sum of absolute empirical Kendall's tau is selected~\cite{dissmann2013selecting}. To illustrate the dependence structure, the first trees of Vine copulas for the all the five categories of YouTube videos are shown in Fig.\ref{fig:five}, along with their bivariate copula families and pairwise empirical Kendall's tau values.

\subsubsection{Copula Families Selection}
After the R-vine structure is selected, the next step is to choose a suitable bivariate copula function for each dependence pair (edge of dependence tree) to fit the data. Due to its quick computation, the independence copula (simple product of two distributions) is first tested to determine if it is suitable to model dependence of two distributions. 

According to Genest~\cite{genest2009goodness}, under the null hypothesis of independence, Kendall's tau is approximately normally distributed with mean of zero. With empirical Kendall's tau value $\tau$, we can choose independence copula using $N$ observations if:
\begin{equation}
\sqrt{\frac{9N(N-1)}{2(2N+5)}}|\tau|<2.
\end{equation}
 If the above independence test fails, then the Akaike Information Criterion (AIC)~\cite{bozdogan1987model} is applied to  choose the desired copula families. By computing the AIC for all possible families, the copula with smallest AIC will be selected:
\begin{equation}
AIC = 2k-2\operatorname{ln}(M).
\end{equation}
where $k$ is the number of estimated parameters of a copula (either 1 or 2 for bivariate copulas), and $M$ is the maximum value among the likelihood functions for all copula candidates~\cite{alexander2008market}. In this work, the bivariate copula families are chosen from 48 possible candidates.
\subsubsection{Parameter Estimation}
Finally, the corresponding parameters for the bivariate copula families $\boldsymbol{\theta}$, which includes $\binom{n}{2}$ parameters for $n$ metrics, can be estimated by using maximum likelihood estimation given data points $\boldsymbol{U}=(x_1, x_2,...x_N)$:
\begin{align}\label{eqn:MLE}
\boldsymbol{\theta^*}& = \operatorname{argmax} L(\boldsymbol{\theta} \vert \boldsymbol{U})  \\ \nonumber
&= \operatorname{argmax} \sum_{i=1}^{N}\sum_{j=1}^{n-1}\sum_{e\in E_j} \operatorname{log} C_{\theta_e}(F(x_{e1}),F(x_{e2})).\\ \nonumber
\end{align}
where $F(x_{e1})$ and $F(x_{e2})$ are the two marginal distributions (nodes of a dependence tree and can be estimated from $\boldsymbol{U}$) connected by copula (edge of a dependence tree) $e$, while $\theta_e$ is the copula parameter for edge $e$. The associated likelihood function is the sum of log-likelihood of all bivariate copulas (i.e all edges $E_j$ of tree $j$), for all $n-1$ trees of a vine over all $N$ observations.

In addition to the regular dependency measure Kendall's tau, the tail dependence measure can also be computed. The upper and lower tail dependence coefficients $\lambda_U$ and $\lambda_L$ are both defined using copula~\cite{embrechts2001modelling}:
\begin{equation}
\label{eqn:hey4}
\operatorname{lim}_{u \rightarrow 1}\frac{1-2u+C(u,u)}{1-u}=\lambda_U,
\end{equation}

\begin{equation*}
\operatorname{lim}_{u \rightarrow 0}\frac{C(u,u)}{u}=\lambda_L.
\end{equation*}

$\lambda_U$ and $\lambda_L$ denote the probabilities that describe the simultaneous co-movements in the upper and lower tails of a bivariate distribution.

\end{document}